\documentclass[12pt,preprint]{aastex}

\newcommand{\Msun}{M_{\sun}}
\newcommand{\Msunsec}{\Msun\,{\mathrm{sec}}^{-1}}
\newcommand{\cm}{{\mathrm{cm}}}

\newcommand{\second}{{\mathrm{s}}}

\newcommand{\gcc}{\mathrm{g}\,\cm^{-3}}

\def\ltaprx {\lower .1ex\hbox{\rlap{\raise .6ex\hbox{\hskip .3ex
        {\ifmmode{\scriptscriptstyle <}\else 
                {$\scriptscriptstyle <$}\fi}}}
        \kern -.4ex{\ifmmode{\scriptscriptstyle \sim}\else 
                {$\scriptscriptstyle\sim$}\fi}}}
\def\gtaprx {\lower .1ex\hbox{\rlap{\raise .6ex\hbox{\hskip .3ex
        {\ifmmode{\scriptscriptstyle >}\else 
                {$\scriptscriptstyle >$}\fi}}}
        \kern -.4ex{\ifmmode{\scriptscriptstyle \sim}\else 
                {$\scriptscriptstyle\sim$}\fi}}}

\shorttitle{Nucleosynthesis in Gamma-Ray Burst Accretion Disks}
\shorttitle{Pruet, Woosley, \& Hoffman}

\begin{document}

\title{NUCLEOSYNTHESIS IN GAMMA-RAY BURST ACCRETION DISKS}
\author{Jason Pruet$^1$, S. E. Woosley$^2$, and R. D. Hoffman$^1$}
\affil{
$^1$N-Division,
    Lawrence Livermore National Laboratory,
    Livermore CA 94550
$^2$Department of Astronomy and Astrophysics, 
    University of California,
    Santa Cruz, CA 95064
}

\begin{abstract}

We follow the nuclear reactions that occur in the accretion disks of
stellar mass black holes that are accreting at a very high rate, 0.01
to 1 $\Msunsec$, as is realized in many current models for gamma-ray
bursts (GRBs). The degree of neutronization in the disk is a sensitive
function of the accretion rate, black hole mass, Kerr parameter, and
disk viscosity. For high accretion rates and low viscosity, material
arriving at the black hole will consist predominantly of
neutrons. This degree of neutronization will have important
implications for the dynamics of the GRB producing jet and perhaps for
the synthesis of the $r$-process. For lower accretion rates and high
viscosity, as might be appropriate for the outer disk in the collapsar
model, neutron-proton equality persists allowing the possible
synthesis of $^{56}$Ni in the disk wind. $^{56}$Ni must be present to
make any optically bright Type I supernova, and in particular those
associated with GRBs.

\end{abstract}


\section{INTRODUCTION}

Growing evidence connects GRBs to the the birth of hyper-accreting
black holes \citep{fry99b}, that is stellar mass black holes accreting
matter from a disk at rates from $\sim$0.01 to 10 $\Msunsec$. Such
models include the collapsar \citep{woo93,mac99}, merging neutron
stars and black holes \citep{eic89,jan99}, supranovae \citep{vie98},
merging helium cores and black holes \citep{zha01}, and merging white
dwarfs and black holes \citep{fry99a}. For such high accretion rates
the disk is optically thick, except to neutrinos, and very hot,
consisting in its inner regions of a (viscous) mixture of neutrons and
protons. At its inner boundary the disk connects to the black hole. 
Whatever processes accelerate the putative GRB-producing jet might
therefore be expected to act upon some mixture of disk material and
other background medium (e.g., the collapsing star in the collapsar
model). Farther out, material will be lost from the disk in a vigorous
wind \citep{mac99,nar02}. In fact, Narayan et al. suggest that {\sl
most} of the disk will be lost to a wind except for those models and
in those regions where neutrino losses dominate the energy budget.

It is thus of some consequence to know the composition of such disks.
In the least case, the nucleosynthesis may be novel and could account
for rare species in nature, such as the $r$-process. At most, the
presence of free neutrons may affect the dynamics of the GRB jet 
\citep{der99,ful00}, the GRB neutrino signature \citep{bah00}, 
light curve \citep{pru02}, and afterglow (e.g. via a ``pre-acceleration''
mechanism similar to the one discussed by Beloborodov 2002).
It is also of some consequence to know
whether the disk wind consists of radioactive $^{56}$Ni, as is
necessary if a visible supernova is to accompany the GRB. If the
electron mole number, $Y_e = \Sigma (Z_i X_i/A_i)$, is less than
0.485, the iron group will be dominated by $^{54,56}$Fe and other more
neutron-rich species \citep{har85} which will be incapable of
illuminating the supernova. Since the jet itself is inefficient at
heating sufficient matter to temperatures required for nuclear
statistical equilibrium (T $\gtaprx 5 \times 10^9$ K), supernovae seen
in conjunction with GRBs \citep[for example]{gal98,blo02} would be
difficult to understand.

We have thus undertaken a survey of the nucleosynthesis that happens
in the disks of rapidly accreting black holes. The work is greatly
facilitated by the existence of numerical and semi-analytic solutions 
that yield the temperature-density structure and drift velocity  
\citep{pop99}. These solutions have been verified in the case of the
collapsar model by direct numerical simulation \citep{mac99}.

\section{COMPUTATIONAL APPROACH}

For a disk that is optically thin to neutrinos the neutron to proton 
ratio, $n/p$, is determined by the competition between electron and
positron capture on nuclei and free nucleons. Since the baryon number
per co-moving volume is conserved, it is convenient to work with the 
electron fraction $Y_e$ defined by
\begin{equation}
\label{yedef}
Y_e = \Sigma \ Z_i (X_i/A_i)
\end{equation} 
where $Z_i$, $X_i$, and $A_i$ are the proton number, mass fraction,
and atomic mass number (integer) of the species $i$. We shall refer to
compositions with $Y_e <$ 0.5 as ``neutron-rich''. 

If we neglect lepton capture on bound nuclei and approximate the disk
material as consisting of a mixture of free nucleons and $\alpha$-particles, 
the evolution of $Y_e$
with radius is given by
\begin{equation}
\label{yedot}
{\bf u}\cdot \left( {\bf \nabla} Y_e\right)=V(r)\sqrt{1-2GM/rc^2 \over 1-(V(r)/c)^2}{dY_e \over dr}=-\lambda_{e^-{\rm p}}\left(Y_e-{1-X_n \over 2}\right) +
\lambda_{e^+{\rm n}}\left(1-Y_e-{1-X_n \over 2}\right).
\end{equation}
Here $X_n$ is the mass fraction of free nucleons, ${\bf u}$ is the
4-velocity of the flow, $M$ represents the black hole mass, and $V(r)$
is the radial drift velocity as measured in an inertial frame
co-rotating with the disk.  In the middle term in Eq. \ref{yedot} we
have assumed steady state conditions as well as cylindrical symmetry, and
we have adopted the relatively simple Schwarzschild metric relevant for most
of our calculations.  Generally, relativistic effects do not make a
big difference in our calculations of the electron fraction, though
they are important for determing the structure of the disk. 

In Eq. \ref{yedot} $\lambda_{ e^-
{\rm p}}$ and $\lambda_{e^+ {\rm n}}$ are the rates for the processes
\begin{eqnarray}
e^-+{\rm p}&\rightarrow& {\rm n}+\nu_e \label{ratefor}\\
e^++{\rm n}&\rightarrow& {\rm p}+\bar \nu_e. 
\label{rateback}\end{eqnarray}
These rates are given by 
\begin{eqnarray}
\lambda_{ e^- {\rm p}}&=&K
	\int_{\delta m \over m_e}^{\infty} w^2(w-{\delta m \over m_e})^2 G_-(1,w)
	S_-(T,U_{\rm F})dw\\
\lambda_{e^+ {\rm n}}&=&K
\int_{0}^{\infty} w^2(w+{\delta m \over m_e})^2 G_+(1,w)
	S_+(T,U_{\rm F})dw.
\end{eqnarray}
Here $K\approx 6.414\cdot 10^{-4} \second^{-1}$ determines the free
neutron lifetime, $\delta m\approx \ 1.293\,{\rm MeV}$ is the neutron
proton mass difference, and $m_e$ is the electron mass.  The functions
$S_-$ and $S_+$ are the electron and positron distribution functions,
while the $G_{\pm}$ are the Coulomb wave correction factors
discussed in \citet{ffn}.

The electron Fermi energy $U_{\rm F}$ 
is found by inverting the expression for the net electron number density,

\begin{equation}
n_e^--n_e^+=\rho Y_e N_A={1\over \pi^2} \left({ m_e c \over \hbar}\right)^3
\int_{0}^{\infty} p^2 (S_-(T,U_{\rm F})-S_+(T,U_{\rm F}))dp
\end{equation}
with $p=\sqrt{w^2-1}$.

We solve for the evolution of $Y_e$ by integrating eq.  \ref{yedot}
for the results presented in \citet{pop99}. Those authors assumed
$Y_e=1/2$. Because we will show that in some cases $Y_e\ll 1/2$, and
consequently that the electron Fermi energy and degeneracy pressure
are substantially smaller than for $Y_e=1/2$, our results are not
entirely self consistent, but should suffice given the very approximate
nature of a one-dimensional calculation.

\section{RESULTS}

\subsection{Scaling with Disk Viscosity and Accretion Rate}

We present results for several different disk models. The parameters
describing these models, as well as the electron fraction near the
event horizon at $r \approx 10^6\cm$, are given in Table
\ref{tbl1}. In all cases we assume $Y_e=1/2$ at large radii. For
simplicity we have adopted a constant black hole mass of 3 $\Msun$
though the scaling relations for mass are obvious in Popham et al. All
else being equal, a larger mass will give less electron capture.

Table \ref{tbl1} shows a clear trend with disk viscosity and accretion
rate: electron capture in the inner disk becomes more pronounced with
decreasing viscosity and increasing mass accretion rate. This arises
because low viscosity flows inefficiently advect angular momentum
outwards and are, for a given mass accretion rate, denser than high
viscosity flows. Dense flows are electron degenerate, so that pair
$e^{\pm}$ creation is suppressed and electron capture dominates over
positron capture. Similarly a larger accretion rate also implies 
a denser disk - for a given value of $\alpha$ - and more capture.

Before discussing in some detail the evolution of $Y_e$ in our
calculations we note that the influence of black hole spin on the
composition is also important. This is because the angular momemtum
imparted to the black hole via the accreting matter will typically
drive the Kerr parameter high. For example, a Kerr parameter of $a\sim
0.9$ is typical for collapsars. A larger Kerr parameter allows the
disk to move to smaller radii before entering the event horizon, so
that more electron capture and a smaller $Y_e$ will result. This is
demonstrated by model G, which corresponds to a black hole with
$a=0.95$. In this model $n/p$ near the event horizon is about ten
times larger than $n/p$ for the same model except with $a=0$ (model
B). The influence of the black hole spin on $Y_e$ is limited to
regions quite near the event horizon. For $r \, > \, 10^{6.5}\cm$, $Y_e$ in
model B is essentially the same as in model G. These considerations
imply that as far as the composition is concerned, the main difference
between flows around zero angular momentum holes and those around
large angular momentum holes will be that the jet in the large angular
momentum case will likely be more neutron rich. The composition in the
bulk of the wind coming from the disk, however, will likely be similar
in the two cases.

\begin{figure}
\epsscale{1.0}
\plotone{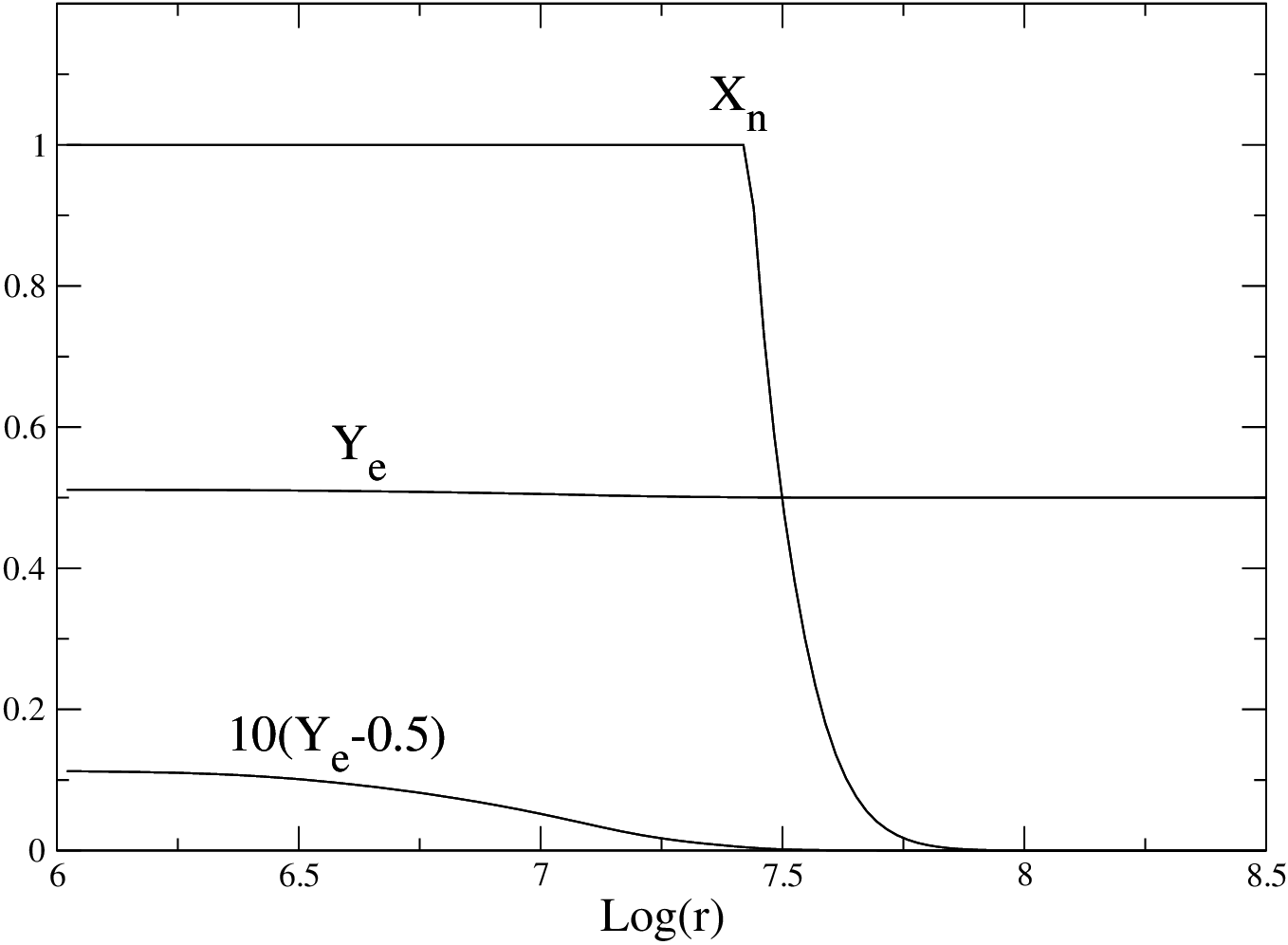}
\caption{Evolution of $Y_e$ and the free nucleon mass fraction 
$X_{n}$ in a disk with $\dot M= 0.01 \Msunsec$ and $\alpha=0.1$.
\label{fig.m.01.alpha.1}}
\end{figure}

In Fig. \ref{fig.m.01.alpha.1} we show the evolution of $Y_e$ with
radius for an accretion disk with $\alpha=0.1$ and a relatively low
mass accretion rate (at least for standard collapsar models), $\dot M
= 0.01 \, \Msunsec$. This flow is not dense enough to drive the electrons
degenerate. Consequently $Y_e$ is governed by thermal $e^{\pm}$
capture. In this case the only asymmetry is the neutron-proton mass
difference and $Y_e$ actually increases slightly owing to the
threshold for the rate in Eq. \ref{ratefor}, becoming greater than
0.5.  The free nucleon mass fraction appearing in
Fig. \ref{fig.m.01.alpha.1} is calculated from the simple estimate
provided in \citet{qia96}. Lepton capture on heavy nuclei is negligible 
for this disk.

Results for a hotter and denser flow, $\dot M=0.1\ \Msunsec$,
$\alpha=0.1$, are shown in Fig. \ref{fig.m1.alpha.1}. The material at
radii $r\lesssim 10^{7.5} \cm$ is mildly degenerate and the electron
fraction is driven to $Y_e\approx 0.44$. The quantity
$\lambda_{e^-{\rm p}} (r/V)$ is the product of the electron capture
rate and a rough measure of the dynamic timescale, or time left before
the infalling material crosses the event horizon. Because $Y_e$ can
not come to equilibrium unless $\lambda_{e^-{\rm p}} (r/V)\approx 1$,
$Y_e$ is not in equilibrium in this flow. Also, because
$\lambda_{e^-{\rm p}} (r/V) \ll 1$ when bound nuclei are present, weak
processes on heavy nuclei are not important even in the extreme limit
where the bound nucleons behave as free nucleons with respect to
electron capture.

\begin{figure}
\epsscale{1.0}
\plotone{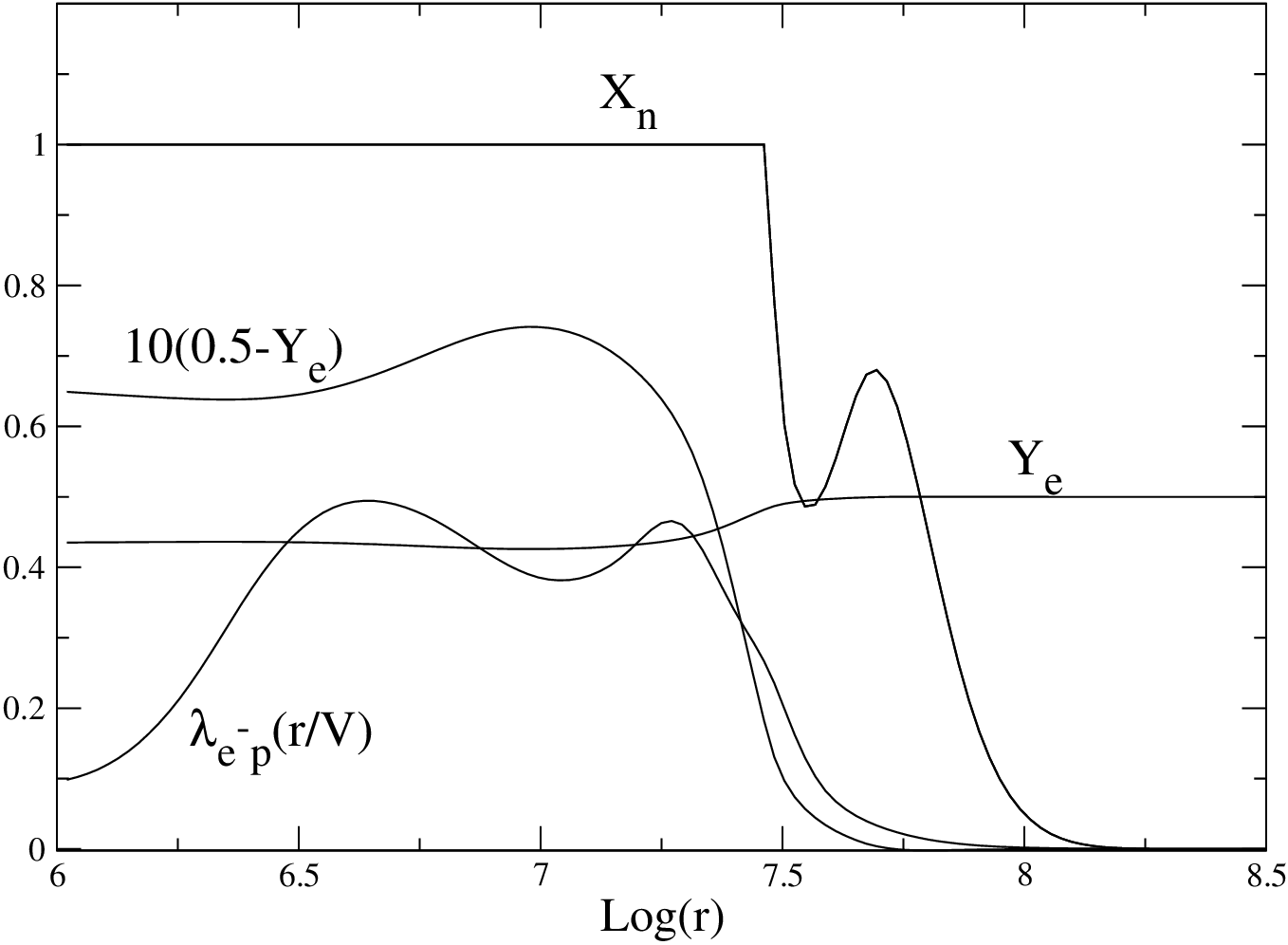}
\caption{Evolution of $Y_e$ and $X_{n}$ in a disk with 
$\dot M=0.1\Msunsec$ and $\alpha=0.1$. Shown also is the quantity 
$\lambda_{e^-{\rm p}}(r/V)$, a rough measure of whether or not there is time
for $Y_e$ to come to equilibrium with respect to $e^{\pm}$ capture.
\label{fig.m.1.alpha.1}}
\end{figure}

\begin{figure}
\epsscale{1.0}
\plotone{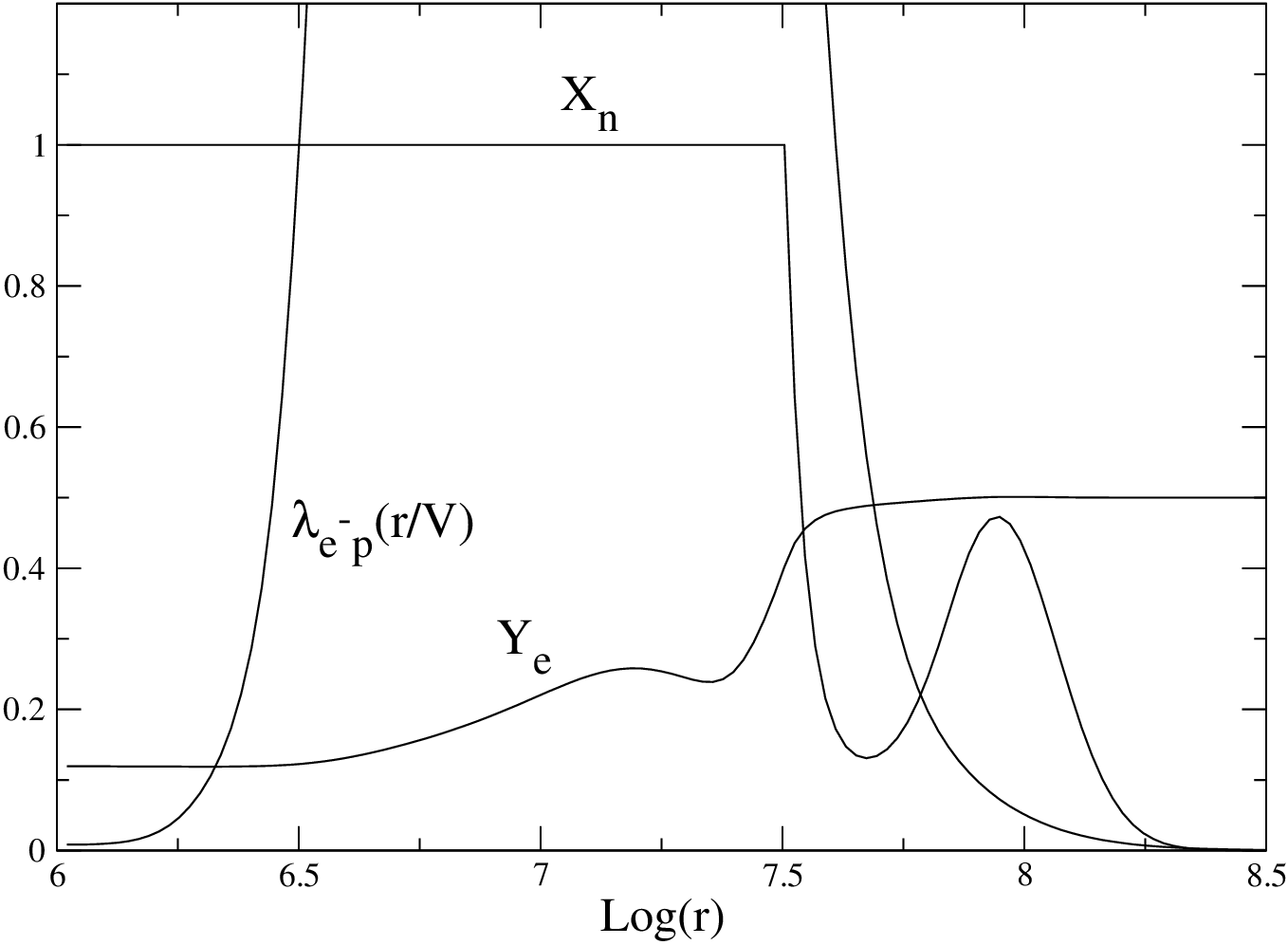}
\caption{Same as Fig. \ref{fig.m.1.alpha.1}, except for $\dot
M=0.1\Msunsec$ and $\alpha=0.03$.  
\label{fig.m.1.alpha.03}}
\end{figure}

\begin{figure}
\epsscale{1.0}
\plotone{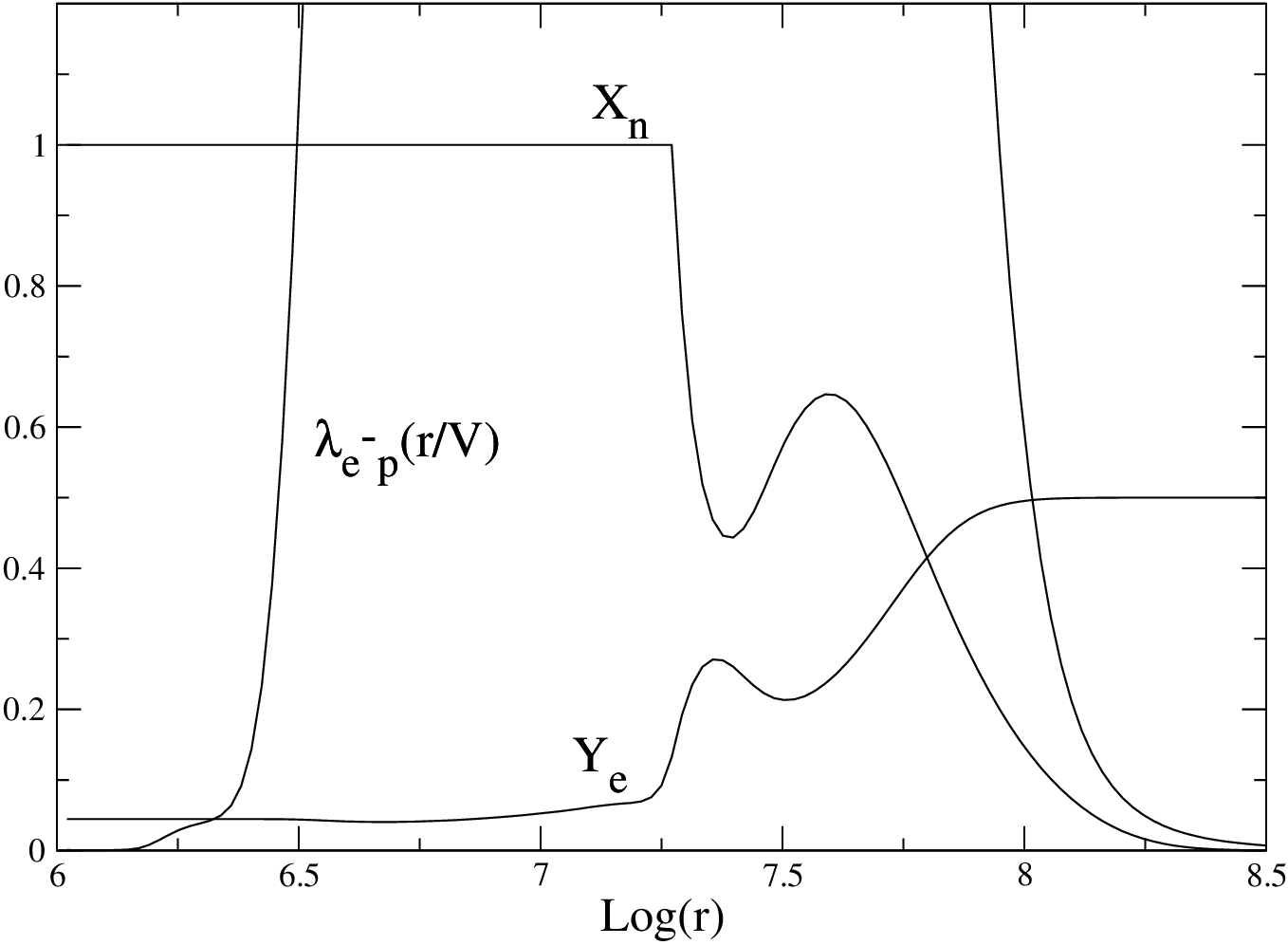}
\caption{Same as Fig. \ref{fig.m.1.alpha.1}, except for $\dot
M=0.1\Msunsec$ and $\alpha=0.01$.  
\label{fig.m.1.alpha.01}}
\end{figure}

\begin{figure}
\epsscale{1.0}
\plotone{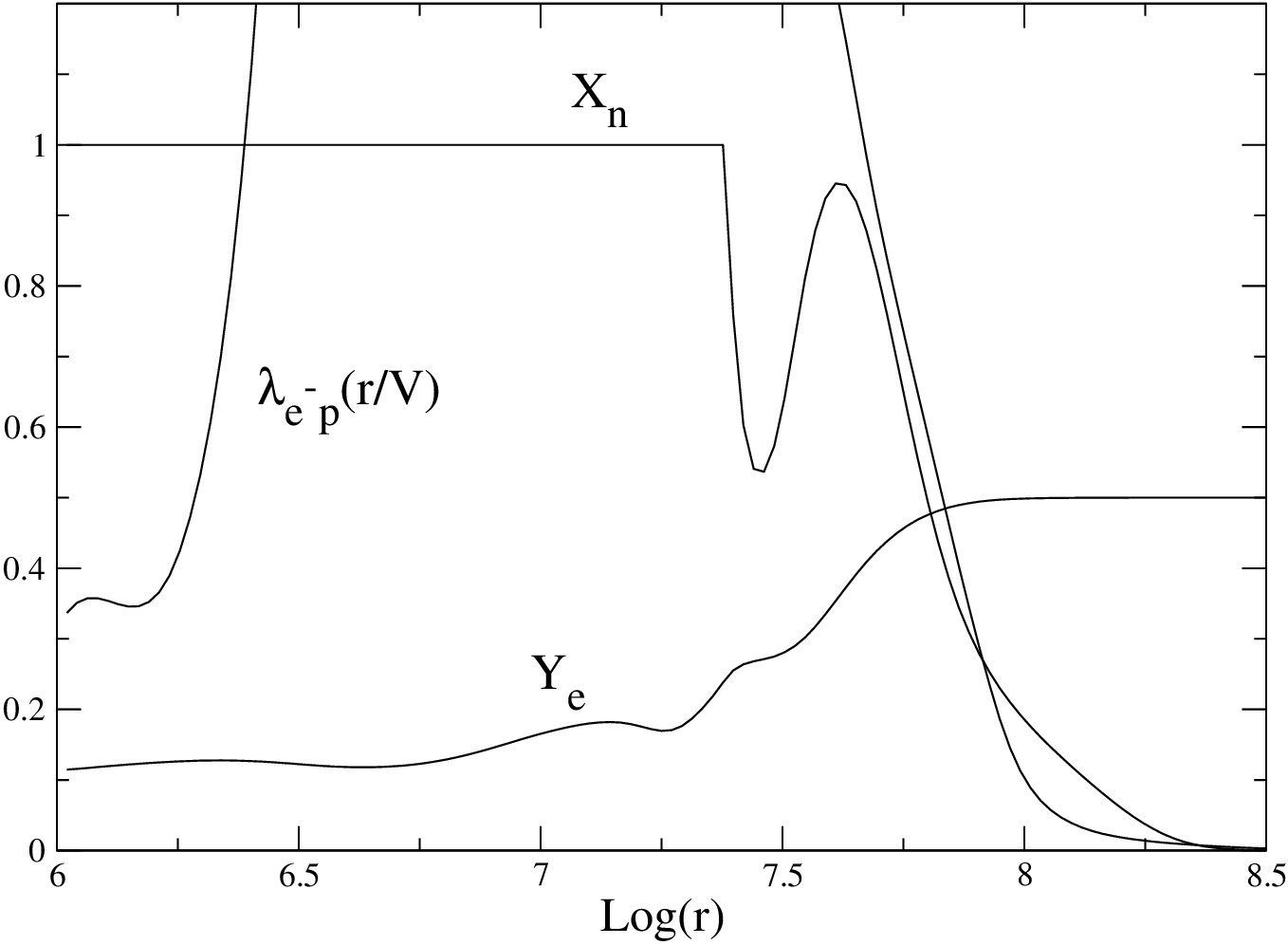}
\caption{Same as Fig. \ref{fig.m.1.alpha.1}, except for $\dot M=1\Msunsec$
and $\alpha=0.1$.
\label{fig.m1.alpha.1}}
\end{figure}

Figures \ref{fig.m.1.alpha.03} and \ref{fig.m.1.alpha.01} show the influence of viscosity on the
composition of the flow. For $\alpha=0.03$ electron capture becomes important
at $r\approx 10^{7.6} \cm$ and $Y_e$ is in close equilibrium with $e^{\pm}$
capture until $r\approx 10^{6.5} \cm$. There is at most $\sim 1/10$  
of an electron capture per bound nucleus in the disk. For $\alpha=0.01$
the inflowing material becomes degenerate and neutron rich early on.
Most of the 
free protons are locked into $\alpha$-particles until the $\alpha$-particles
dissociate. Positron capture on the excess free neutrons results in a 
brief {\it increase} in $Y_e$, as seen in the bump at $r\approx 10^{7.3}\cm$
in Fig. \ref{fig.m.1.alpha.1}. The final $n/p$ in this disk is very large,
$\sim 20$.

The influence of the mass accretion rate on the composition is seen in
Fig. \ref{fig.m1.alpha.1}, where we plot results for $\dot M = 1\
\Msunsec$ and $\alpha=0.1$. This disk is quite similar to the $\dot M=
0.1 \ \Msunsec$, $\alpha =0.3$ case. Indeed, this is just what is
expected from the scaling relations in \citet{pop99}.  For $\dot
M\approx 1\ \Msunsec$ and larger, the assumption that the disk is
optically thin to neutrinos begins to break down \citep{mac99,dim02}. This
influences both the structure of the disk, and, because neutrino
capture becomes important, the evolution of $Y_e$. Because neutrino
trapping likely influences the structure of the disk at the same level
as the assumption that $Y_e = 1/2$, we do not calculate $Y_e$ for disks
calculated under the assumption of partial neutrino trapping (and
$Y_e = 1/2$).

\subsection{Effects of Neutrino Capture}

The effect of neutrino capture on the evolution of $Y_e$ can 
be addressed in an approximate way. In particular, we are interested 
in the rate $\lambda_{\nu_e {\rm n}}$ for the process 
\begin{equation}
\label{nuen}
\nu_e+n\rightarrow {\rm p} + e^-.
\end{equation} 
The rate for $\bar \nu_e$ capture is suppressed both because of the
low proton number density and the absence of high energy positrons in the
flow. The number of neutrinos captured per neutron is roughly 
$\tau_{\nu_e,{\rm cap}} n_{\nu_e}/n_{\rm n} \approx Y_e 
\tau_{\nu_e,{\rm cap}} n_{\nu_e}/n_{\rm p}$, 
where $\tau_{\nu_e,{\rm cap}}$ is the optical depth for the 
process in Eq. \ref{nuen}, and $n_{\nu_e}/n_{\rm n}$ ($n_{\nu_e}/n_{\rm p}$)
is the number of electron neutrinos produced per neutron (proton) in the flow.
The ratio $R$ of the number of neutrino captures per neutron to the 
number of electron captures per proton is 
\begin{equation}
\label{effectratio}
R \approx {\lambda_{\nu_e}\over \lambda_{e^-{\rm p}}} =
\tau_{\nu_e,{\rm cap}} Y_e.
\end{equation}
This equation implies that an optical depth of 1 to $\nu_e$ capture is
roughly equivalent to a doubling of $\lambda_{e^+{\rm n}}$. We note here
that while the the $\nu_e$'s produced in electron-capture reactions
are not thermal, their average capture cross section is only
approximately a factor of 2 higher than the average capture cross
section calculated under the assumption of a thermal distribution for
the electron neutrinos (at the electron Fermi energy and temperature)
for conditions of interest in accretion disks.

It is difficult to quantitatively calculate the influence of neutrino
capture on $Y_e$ in a post-processing step. This is both because of
the complexities of neutrino transport, as well as because of the
feedback between $Y_e$, the electron capture rate, and the disk
dynamics. However, the following simple considerations argue that
neutrino capture is unlikely to have a dramatic influence on $Y_e$.

To discuss some
specific cases, consider first the $\dot M=0.1\Msunsec$, $\alpha=0.1$
disk. \citet{dim02} argue that neutrinos will have an absorptive optical
depth of unity at $r\approx 3\cdot 10^{6}\cm$ for this disk. Because
weak processes have essentially frozen out by this radius (see the
$\lambda_{e^-{\rm p}}(r/V)$ curve in Fig. \ref{fig.m.1.alpha.1}), neutrino
capture will have a small influence on $Y_e$. 

Denser flows trap neutrinos more efficiently but are still expected to
remain neutron rich. For example, for the $\dot M=1\Msunsec$,
$\alpha=0.1$ disk, \citet{dim02} show that the neutrino absorptive
optical depth will be greater than unity for $r\lesssim 2\cdot
10^7\cm$. As weak processes are rapid compared to the dynamic
timescale in this disk, the result will be an increase in $Y_e$.  To
estimate the magnitude of this increase we adopt the simple procedure
of artificially increasing the positron capture rate,
$\lambda_{e^+{\rm n}}\rightarrow 5\lambda_{e^+{\rm n}}$, which is
likely an overestimate of the protonization rate. The increase in
$\lambda_{e^+{\rm n}}$results in an electron fraction of $Y_e\approx
0.166$ near the event horizon. This is approximately 50\% higher than
the electron fraction calculated by neglecting the influence of
$\nu_e$ capture, though still quite neutron rich.

To summarize, $Y_e$ in disks with high viscosity and modest accretion
rates will not be affected by neutrino capture simply because
neutrinos are not trapped or because weak processes are too slow. The
electron fraction in disks with neutrino absorptive optical depths of
a few will increase somewhat but will remain neutron rich owing to the
sharp rise in $\lambda_{e^-{\rm p}}$ with $Y_e$.

\subsection{Transport to the Surface of the Disk}

In order for a low electron fraction to have observable 
implications the electron fraction must not come to equilibrium
at $Y_e=1/2$ as the material travels out of the plane of the disk. To estimate
the evolution of the electron fraction in convective blobs moving out
of the disk we parametrize the convective timescale by the 
turbulent convection speed \citep{nar02} 
\begin{equation}
v_{\rm turb} \approx \alpha v_K = 6.3 \times
10^8 \alpha_{-1} r_7^{-1/2} \ {\rm  cm \ s^{-1}}
\end{equation}
where $\alpha_{-1}=\alpha/0.1$ and $r_7=r/10^7 \cm$. 
This implies a time to go one pressure scale height 
\begin{equation}
\tau_{\rm conv} \sim r/v_{\rm turb} \sim 16 \ \alpha_{-1}^{-1} \, r_7^{1/2}
\ {\rm ms}.
\end{equation}
We also assume that the convective blob expands adiabatically. 
An estimate for the entropy per baryon in the disk is given by
\citet{qia96}:
\begin{equation}
\label{entropy}
s/k_b\approx 0.052 {T_{\rm MeV}^3 \over \rho_{10}}+
7.4+\ln\left({T^{3/2}_{\rm MeV}\over \rho_{10}}\right).
\end{equation}
Here $T_{\rm MeV}$ is the temperature in MeV and $\rho_{10}$ is the
density in units of $10^{10} \gcc$.  The first term on the right hand
side of Eq. \ref{entropy} is the contribution to the entropy from
relativistic light particles ($\gamma/e^{\pm}$), while the next two
represent the contribution from free nucleons.  Eq. \ref{entropy} is
not appropriate when electrons are degenerate. For degenerate
electrons the contribution of thermal $e^{\pm}$ pairs is small and the
coefficient 0.052 should be closer to 0.02, representing just the
entropy of the photons. However, changing the coefficient to 0.02
makes little difference for our purposes.

When relativistic particles dominate the entropy, the adiabat satisfies
$T^3\propto \rho$. When free nucleons dominate the entropy, 
the adiabat satisfies $T^{3/2}\propto\rho$. The evolution of the electron
fraction in a convective blob is crucially sensitive to the adiabat 
on which the blob travels.

\begin{figure}
\epsscale{1.0}
\plotone{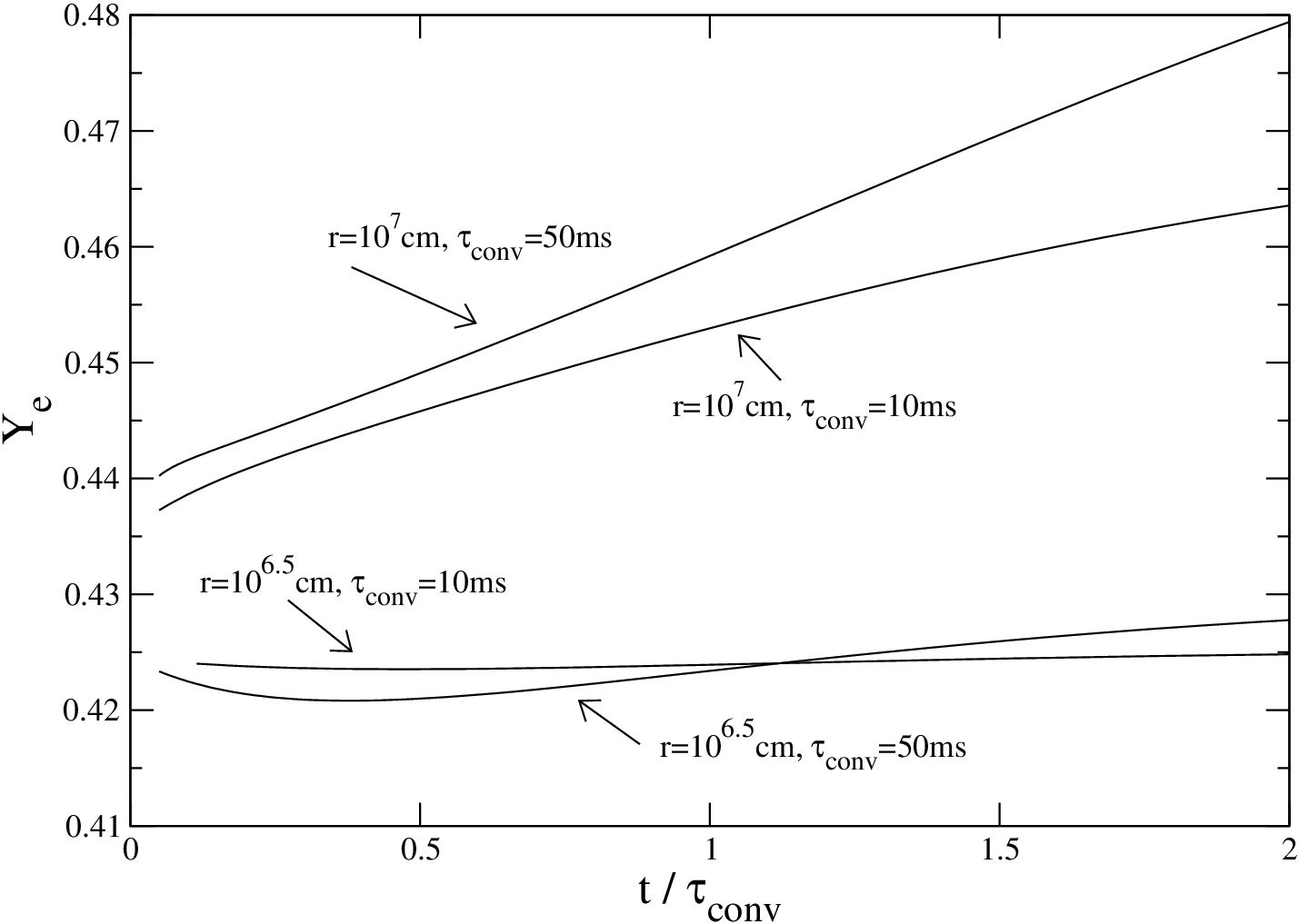}
\caption{Evolution of electron fraction in a convective and adiabatic
fluid element in a disk with $\dot M=0.1\Msunsec$ and $\alpha=0.1$. 
The assumed convective timescale $\tau_{\rm conv}$ and radius in the disk
from which the material originates is given next to each curve.
\label{blob1}}
\end{figure}

To illustrate this, consider first a disk with $\dot M=0.01$,
$\alpha=0.1$.  In Fig. \ref{blob1} we show the evolution of the
electron fraction in matter originating from two different points in
this disk and for different assumptions about $\tau_{\rm conv}$. The
upper curves in Fig. \ref{blob1} correspond to material originating
from $r=10^{6.5}\cm$ in the disk. At this point the entropy is
relatively high, $s/k_b\sim 23$ by the estimate in Eq. \ref{entropy},
and consequently the adiabat is closer to $T^3\propto \rho$ than to
$T^{3/2}\propto\rho$.  The temperature decreases quite slowly relative
to the density, allowing for the formation of pairs which efficiently
drive $n/p$ to equality. For large $\tau_{\rm conv}$ this process
occurs more efficiently than for small $\tau_{\rm conv}$.  The lower
curves in Fig. \ref{blob1} correspond to matter originating from
$r=10^7\cm$ in the disk. At this point the entropy is $\approx 17$, so
the adiabat is near $T^{3/2}\propto \rho$. For low entropies, then,
the temperature and density decrease at a comparable rate and pair
formation is somewhat suppressed, keeping $Y_e$ low.

\begin{figure}
\epsscale{1.0}
\plotone{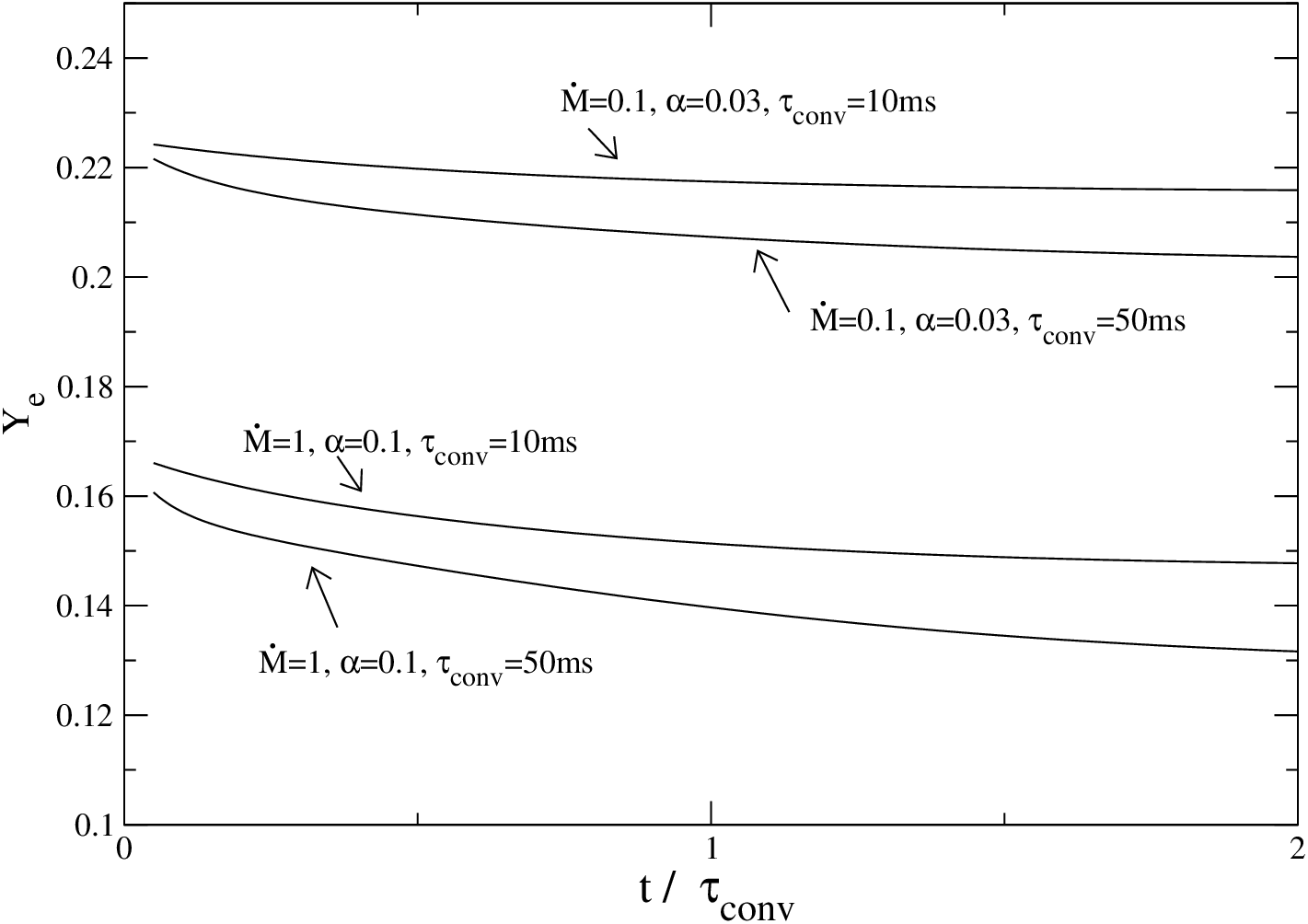}
\caption{Evolution of electron fraction in a convective and adiabatic
fluid element in two different disks characterized by low electron fractions.
The assumed convective timescale and disk parameters are given next to each curve.
In all cases the material is assumed to originate from $r=10^7\cm$. 
\label{blob2}}
\end{figure}

For flows where the electron fraction is high enough to drive the
inner disk neutron rich, the entropy is generally dominated by
non-relativistic particles, and the adiabat again preserves a low
$Y_e$. This is illustrated in Fig. \ref{blob2} where we plot the
evolution of electron fraction in an adiabatic convective bubble for
two different disks characterized by a low $Y_e$. In this figure we
only show the evolution of a fluid element from a single initial
radius ($10^7\cm$) because the entropy in these is disks never
dominated by relativistic particles when $Y_e$ is low, so that the
behavior shown in Fig. \ref{blob2} is generic.

The above considerations outline the general features of how $Y_e$
evolves in material as it travels from the center to the surface of
the disk. In flows where $Y_e$ is driven small, the degeneracy of the
electrons implies that free nucleons dominate the entropy. A low $Y_e$
then, will remain low.  By contrast, in flows where $Y_e$ is close to
1/2, the adiabat can be close enough to $T^3\propto \rho$ for $Y_e$ to
be driven to 1/2, and the neutron excess at the center of the disk is
larger than the neutron excess in material finding its way out of the
disk.

%
%
%

\section{IMPLICATIONS}

\subsection{Radioactivity in the disk wind}

The most obvious consequence of our calculations is that $^{56}$Ni
will be absent from the winds of hyper-accreting black holes unless a)
the accretion rate is low ($\ltaprx 0.1\ \Msunsec$), and b) the disk
viscosity is high, $\alpha \gtaprx 0.1$. Interestingly, modern views
regarding $\alpha$-disks and GRB models favor values close to these.

Typical accretion rates in the collapsar model are 0.05 - 0.1 
$\Msunsec$ \citep[their figures 5 and 10]{mac99}.  The neutron excess
will be smaller in Type II collapsars powered by fall back rather
than direct black hole formation \citep{mac01}. Considerable
accretion into the hole and mass loss from the disk may continue, at a
declining rate, even after the main GRB producing event ($\sim$20 s)
is over \citep{zha02}. It thus seems likely that the collapsar model
will be able to provide the $^{56}$Ni necessary to make the supernovae
that accompany GRBs (though only if $\alpha \gtaprx 0.1$). This
is also true of the slower accreting models like helium-star black
hole mergers and black hole white dwarf mergers. However any wind from 
merging compact objects, or similar models like the supranova, will be
neutron rich. Though perhaps of interest for nucleosynthesis, they
will not produce $^{56}$Ni, at least during the black hole accretion
epoch.

\subsection{Neutron excess in GRB jets}

For disks with high accretion rates, and certainly for merging neutron
stars or black hole neutron star mergers, the matter near the event
horizon will be very neutron rich. If this material pollutes the
outgoing jet, the the GRB jet will itself, at least initially, contain
free neutrons. 

The dynamics of accelerating neutron-rich jets can differ
dramatically from the dynamics of pure proton jets. \citet{ful00} 
showed that a high Lorentz factor fireball that is neutron rich 
can lead to two very distinct kinematic components, a slow neutron outflow
and a fast proton outflow. This arises because the uncharged neutrons
are weakly coupled to the radiation dominated plasma, and are
accelerated principally via strong neutron-proton scatterings \citep{der99}.
Strong scatterings freeze out when they become slow compared to the dynamic
timescale and at this point the neutrons coast. If this decoupling
occurs while the jet is still accelerating, then the coulomb-coupled 
protons go on to have a larger Lorentz factor than the neutrons. 

Roughly, dynamic neutron decoupling is only expected for fast jets. Here 
the
precise meaning of ``fast'' depends, among other things, the fireball
source size. For relativistic flows originating from compact objects, the 
dynamic timescale characterizing the acceleration of the flow is about 1 ms
and the final Lorentz factor must be greater than $\sim300$ in order for 
dynamic neutron decoupling to occur. For jets in the collapsar model,
the timescale characterizing the acceleration is set by the surface of last 
interaction of the jet with the stellar envelope at $\sim10^{11}\cm$. In this 
case the final Lorentz factor of the jet has to be $\gtrsim3000$ for dynamic
neutron decoupling to occur. Because this is an impossibly high Lorentz factor
for the collapsar model, dynamic neutron decoupling will not occur.

There may be a detectable neutrino signature of dynamic neutron
decoupling.  Neutron-proton collisions occurring during decoupling will
generate pions.  In turn these pions will decay and lead to the
generation of neutrinos with energies of a few GeV in our reference
frame. These neutrinos should be detectable at the rate of about one
per year in next generation neutrino telescopes \citep{bah00}. In
addition, there may also be a direct electromagnetic signature of
neutron decoupling.  For some bursts arising from external shocks, a
slow decoupled neutron shell will decay and shock with the outer
proton shell as the outer shell plows into the interstellar
medium. This leads to a characteristic two-peaked structure in the
burst \citep{pru02}.

Regardless of whether dynamic neutron decoupling occurs, neutrons in
the jet may have observable implications for the GRB afterglow. This
is particularly true when the shocking radius, i.e. the radius at which the
shocks giving rise to the observed $\gamma-$rays occur, is smaller
than the length over which free neutrons decay. This condition is
\begin{equation}
\label{nafterglow}
r_{\rm shock}\lesssim \gamma_2 \tau_n c \approx 10^{15}\gamma_2 \cm,
\end{equation}
where $r_{\rm shock}$ is the shocking radius, $\gamma_2$ is the 
Lorentz factor of the outflow in units of 100, and $\tau_n\sim 1000\sec$ is
the free neutron lifetime. Eq. \ref{nafterglow} is generally satisfied
for bursts from internal shocks and bursts from external shocks in a very dense
medium. When Eq. \ref{nafterglow} holds, the neutrons can stream
ahead of the slowing proton shell (the complement of the dynamic neutron 
decoupling discussed above) and deposit energy as they decay.
Implications for the resulting afterglow in such a case have been discussed
by \citet{bel02}.

In the next section we discuss the r-process in winds and jets from
accretion disks. However, we first note that GRB jets are characterized
by interesting light element synthesis. \citet{pgf02} and \citet{lem02} 
calculated the thermal synthesis of light elements in GRB-like outflows. 
They find that the final deuterium mass fraction can be of order 1\%,
some three orders of magnitude larger than the primordial yield. For 
kinematically well coupled flows, the nucleosynthesis depends sensitively
on $Y_e$, with deuterium yield decreasing with increasing neutron
excess. For flows in which dynamic neutron decoupling occurs, high energy 
neutron-$\alpha$ collisions will spall deuterons and can result in final 
deuteron mass fractions as high as 10\% \citep{pgf02}.

\subsection{The r-process} 

There are two possible sites for the $r$-process here in those cases
where $Y_e$ is low - in the jet and in the disk wind.

\subsubsection{In the jet}

The jet has the merit of originating in the vicinity of the black hole
where the neutron excess is likely to be greatest. If magnetic fields
drive the heating and initial acceleration of the jet, then this
neutron excess will likely be preserved. If neutrinos drive the
outflow \citep{mac99}, $Y_e$ will be reset to some extent by the weak
interactions. However, if the details of the outflow
above the black hole are similar to spherically symmetric neutrino
driven ultra-relativistic winds, then only $\nu \bar \nu$ annihilation
will be important and the flow will remain neutron rich \citep{pru01}.

It is clear that the rapid expansion of the jet will be favorable to
freezing out with a large abundance of free neutrons and this is
conducive to the $r$-process \citep{hof97}. However, a pure nucleonic
jet coming from the adiabatic expansion of a fireball with energy
loading $\eta = E_{\rm internal}/\rho c^2 \gtaprx 100$ is too much of
a good thing when it comes to entropy and rapid expansion. The entropy
was given in eq. \ref{entropy} \citep{qia96}
\begin{equation}
s/k = \frac{11 \pi^2}{45}(\frac{kT}{\hbar c})^3{m_N/\rho}  
\end{equation}
The internal energy (erg/gm) is
\begin{equation}
\epsilon = \frac{11 \pi^2}{60}(\frac{kT}{\hbar c})^4{1/\rho}  
\end{equation}
and hence, for $\eta = 200$ and $kT \sim$ 2 MeV
\begin{equation}
s/k_b = \frac{4}{3} \frac{\eta c^2 m_{\rm N}}{kT} \sim 10^5.
\end{equation}

The time scale for the expansion, $r/c$, is less than a millisecond. 
The high entropy implies a low density at the temperature when $n$ and
$p$ can start to recombine. Coupled with the rapid expansion, one
might conclude that the jet will remain pure nucleons and light nuclei 
and that an
$r$-process is impossible.

This conclusion would be erroneous on several counts. First, the jet
may not always have such high energy loading. There may be cases where
no gamma-ray burst is produced and the energy loading of the jet is
much less. The entropy would then decrease and the expansion time
scale would increase. Whether to call these ``dirty fireballs'' or
simply an extension of the ``disk wind'' is simply a matter of taste
(see below).

Of more relevance to GRB models, and particularly to those that
involve massive stars, is that the jet will not escape the star
without interaction. Along the walls of the jet, the Kelvin Helmholtz
instability will occur \citep{zha02}. The jet will also encounter one
or more shocks where it will be abruptly slowed. Most interesting is
the ``jet head'' which moves through the star subrelativistically with
a speed that varies but is of order c/3. At this head the jet mixes
with stellar material and the mixture is swept backwards (in the
moving frame) forming a cocoon. It typically takes 5 to 10 sec for the 
jet head to penetrate the roughly one solar radius of overlying
Wolf-Rayet star in the collapsar model.

During this time, neutrons in the jet are mixed with heavy nuclei -
He, O, Si, Ne, Mg and the like - which can serve as seeds for an
$r$-process. The overall dynamics is likely to be quite complex and
its study would requires a multi-dimensional relativistic simulation
well beyond the scope of this paper. However, \cite{zha02} find
typical densities and time scales in the jet prior to break out are
10$^{-4}$ to 10$^{-1}$ g cm$^{-3}$ and seconds. This is enough that
many, if not all of the neutrons in the jet would capture.

Still, the overall yield in gamma-ray bursts is too small to account
for the solar $r$-process. Assume there is one GRB for each 100
supernovae. Each burst has 10$^{51}$ - 10$^{52}$ erg of relativistic
ejecta (both jets) and a Lorentz factor that is at least 200
\citep{lit01}. The jet will be some mixture of entrained stellar
material and jet, but assume that the initial jet is half or more of
the relativistic ejecta. This gives an equivalent yield of about
10$^{-7}$ $\Msun$ per supernova. Probably the $r$-process will be at
most 10\% of this (the rest may be $\alpha$-particles and entrained
nuclei). This then is several orders of magnitude less than required
to make the solar r-process. However, GRB jets might still be important
sources of the r-process in metal-deficient stars or of select rare
nuclei in the sun.

\subsubsection{The $r$-process in the disk wind}

A more significant $r$-process could come from the non-relativistic
ejecta that comprise the disk wind, provided that the entropy is
increased by magnetic processes to well above its value in the middle
of the disk, as in for example, \citet{dai02}.  Assume that the
accretion is 50\% efficient, that is that half the material that
enters the disk eventually enters the black hole and the other half is
blown away \citep{nar02}. A typical collapsar powered GRB involves the
accretion of from one to several solar masses, so $\sim$1 $\Msun$ is
ejected from the disk. Most of this will not be neutron-rich enough or
expand fast enough to make the $r$-process, but consider the
implications if only 1\% of the mass that is lost comes from the inner
disk at times when the accretion rate is over 0.1 $\Msunsec$ and $Y_e
< 0.4$.

So close to the hole, significant entropy will be added by any
acceleration process that produces a strong outflow. A crude estimate is
$s/k_b \sim G M m_n/(r k T_{max}) \sim 400 \, r_6^{-1} (2 \ {\rm
MeV}/T_{max})$ \citep{qia96}, where $T_{max}$ is the temperature where
the energy deposition is the greatest. In fact the conditions in the
inner disk for accretion rates $\sim$0.1 $\Msunsec$ are not so
different - in terms of neutrino luminosity, neutron excess,
temperature, and gravitational potential - from the neutrino-driven
wind in an ordinary supernova, and one might expect a similar
$r$-process \citep{woo94}.  The gravitational potential though, can in
principle, be greater and a larger entropy and shorter time
scale might be realized.  Assume that the ejecta will be very rich in
$\alpha$-particles and contain of order 10\% $r$-process by
mass. Putting the numbers together then gives an $r$-process synthesis
equivalent to 10$^{-5}$ $\Msun$ per supernova, making this possibility
well worth further investigation.

\acknowledgments

The authors acknowledge helpful correspondence with Weiqun Zhang
regarding the conditions in collapsar jets. 
This research has been supported by NASA (NAG5-8128, NAG5-12036, and
MIT-292701) and the DOE Program for Scientific Discovery through
Advanced Computing (SciDAC; DE-FC02-01ER41176). A portion of this work 
was performed under the auspices of the U.S. Department of Energy
by University of California Lawrence Livermore Laboratory under 
contract W-7405-ENG-48.

\clearpage

\begin{deluxetable}{ccccc}
\tablecaption{Electron Mole Number Near the Event Horizon \label{tbl1}}
\tablewidth{0pt}
\tablehead{
\colhead{Model} &
\colhead{$\dot M$\tablenotemark{a}} &
\colhead{$\alpha$\tablenotemark{b}} &
\colhead{$Y_e$} }
\startdata
 A & 0.01 & 0.1  & 0.510 \\
 B & 0.1  & 0.1  & 0.435 \\
 C & 0.1  & 0.03 & 0.119 \\ 
 D & 0.1  & 0.01 & 0.045 \\
 E & 1.0  & 0.1  & 0.115 \\ 
 F\tablenotemark{c} & 0.03  & 0.1 & 0.527 \\
 G\tablenotemark{d} & 0.1 & 0.1 &\ \ \ \  0.077 
\tablenotetext{a}{Accretion rate in $\Msunsec$}
\tablenotetext{b}{Disk viscosity}
\tablenotetext{c}{Disk properties inferred from the scaling relations
in \cite{pop99} for this model.}
\tablenotetext{d}{Kerr parameter $a=0.95$ for this model.}

\enddata
\end{deluxetable}

\clearpage

\clearpage

\clearpage

\clearpage

\clearpage

\clearpage

\clearpage


\begin{thebibliography}{}

\bibitem[Bahcall \& M\'esz\'aros(2000)]{bah00}
Bahcall, J.~N. \& M\'esz\'aros, P. 2000, \prl 85, 1362.

\bibitem[Beloborodov(2002)]{bel02}
Beloborodov, A. M. 2002, astro-ph/0209228.


\bibitem[Bloom et al.(2002)]{blo02} 
Bloom, J. S., Kulkarni, S. R., Price, P. A., Reichart, D., Galama,
T. J., Schmidt, B. P., Frail, D. A., Berger, E., et al 2002, \apjl,
572, L45

\bibitem[Daigne \& Mochkovitch (2002)]{dai02}
Daigne, F., \& Mochkovitch, R. 2002 \aap, 388, 189

\bibitem[Derishev, Kocharovsky, \& Kocharovsky(1999)]{der99} Derishev,
E.~V., Kocharovsky, V.~V., \& Kocharovsky, Vl.V. 1999, A\&A, 345, 51.

\bibitem[Di Matteo , Perna, \& Narayan (2002)]{dim02}
Di Matteo, T., Perna, R., \& Narayan, R. 2002 \aap, 390, L35 (astroph 0207319) 

\bibitem[Eichler et al.(1989)]{eic89}
Eichler, D., Livio, M., Piran, T., \& Schramm, D. N. 
1989, Nature, 340, 126 


\bibitem[Frail et al.(2001)]{fra01}
Frail, D., et al. 2001, \apjl, 562, L55

\bibitem[Freedman \& Waxman(2001)]{fre01}
Freedman, D. L., \& Waxman, E. 2001, \apj, 547, 922

\bibitem[Fryer et al. (1999)]{fry99a}
Fryer, C. L., Woosley, S. E., Herant, M., \& Davies, M. B. 1999
\apj, 520, 650

\bibitem[Fryer, Woosley, \& Hartmann (1999)]{fry99b}
Fryer, C. L., Woosley, S. E., \& Hartmann, D. H. 1999, \apj, 526, 152

\bibitem[Fuller, Fowler, \& Newman(1982)]{ffn}
Fuller, G. M., Fowler, W. A., \& Newman, M. J. 1982, \apj, 252, 715.

\bibitem[Fuller, Pruet, \& Abazajian(2000)]{ful00} Fuller, G.~M.,
Pruet, J., \& Abazajian, K. 2000, \prl, 85, 2673.

\bibitem[Galama et al. (1998)]{gal98}
Galama, T. J., Vreeswijk, P. M., van Paradijs, J., Kouveliotou, C.,
Augusteijn, T., Bohnhardt, H., Brewer, J. P., Doublier, V., et al. 1998,
Nature, 395, 670

\bibitem[Hartmann, Woosley, \& El Eid (1985)]{har85}
Hartmann, D., Woosley, S. E., \& El Eid, M. 1985 \apj, 297, 837
97ApJ...482..951H

\bibitem[Hoffman, Woosley, \& Qian (1997)]{hof97}
Hoffman, R. D., Woosley, S. E., \& Qian, Y.-Z. 1997, \apj, 482, 951

\bibitem[Janka et al.(1999, 2001)]{jan99}
Janka, H.-T., Eberl, T., Ruffert, M., \& Fryer, C. L. 1999 \apj, 527, 39

\bibitem[Lemoine(2002)]{lem02}
Lemoine, M. 2002, \aap\ \ letters, in press. 

\bibitem[Lithwick \& Sari (2001)]{lit01}
Lithwick, Y., \& Sari, R. 2001, \apj, 555, 540

\bibitem[MacFadyen \& Woosley(1999)]{mac99}
MacFadyen, A. I., \& Woosley, S. E. 1999, \apj, 524, 262

\bibitem[MacFadyen, Woosley, \& Heger(2001)]{mac01}
MacFadyen, A. I., Woosley, S. E., \& Heger, A. 2001, \apj, 550, 410

\bibitem[Narayan, Piran, \& Kumar (2002)]{nar02}
Narayan, R., Piran, T., Kumar, P. 2001 \apj, 557, 949

\bibitem[Popham, Woosley, \& Fryer (1999)]{pop99}
Popham, R., Woosley, S. E., \& Fryer C. L 1999, \apj, 
518, 356

\bibitem[Pruet \& Dalal (2002)]{pru02}
Pruet, J., \& Dalal, N. 2002, \apj, 573, 770


\bibitem[Pruet, Fuller, \& Cardall(2001)]{pru01}
Pruet, J., Fuller, G. M., \& Cardall, C.Y. 2001, \apj, 561, 957


\bibitem[Pruet, Guiles, \& Fuller(2002)]{pgf02}
Pruet, J., Guiles, S., \& Fuller 2002, \apj, in press.

\bibitem[Qian \& Woosley (1996)]{qia96}
Qian, Y.-Z., \& Woosley, S. E. 1996, \apj, 471, 331

\bibitem[Ruffert \& Janka (2001)]{ruf01}
Ruffert, M., \& Janka, H.-Th. 2001, \aap, 380, 544

\bibitem[Vietri \& Stella (1998, 1999)]{vie98}
Vietri, M., \& Stella, L. 1998, \apjl, 507, L45 

\bibitem[Vietri \& Stella (1999)]{vie99}
Vietri, M., \& Stella, L. 1999, \apjl, 527, L43 

\bibitem[Woosley, Arnett, \& Clayton (1973)]{woo73}
Woosley, S. E., Arnett, W. D., \&  Clayton, D. D. 1973, 
\apjs, 26, 231

\bibitem[Woosley (1993)]{woo93}
Woosley, S. E. 1993, \apj, 405, 273

\bibitem[Woosley et al. (1994)]{woo94}
Woosley, S. E., Wilson, J. R., Mathews, G. J., Hoffman, R. D.,
\& Meyer, B. S. 1994, \apj, 433, 229

\bibitem[Zhang \& Fryer (2001)]{zha01}
Zhang, W., \& Fryer, C. L 2001 \apj, 550, 357

\bibitem[Zhang \& Woosley (2002)]{zha02}
Zhang, W. \& Woosley, S. E. 2002, \apj, submitted, 
astro-ph/0207436 


\end{thebibliography}
\end{document}